\documentclass[aps,physrev,reprint,superscriptaddress,longbibliography]{revtex4-2}

\usepackage{graphicx}
\usepackage{dcolumn}
\usepackage{bm}

\usepackage[colorlinks=true,allcolors=blue]{hyperref}
\usepackage{xurl}
\usepackage{amsmath}
\usepackage{mathtools}
\usepackage{amssymb}
\usepackage{bm}
\usepackage[normalem]{ulem}
\usepackage{microtype}
\usepackage{subcaption}
\usepackage{soul}
\usepackage{bbm}
\usepackage{bbold}
\usepackage{multirow}
\usepackage{array}
\usepackage{stackengine}
\usepackage{braket}
\usepackage{comment}

\renewcommand{\vec}[1]{\boldsymbol{#1}}

\begin{document}

\title{Multipartite entanglement vs nonlocality for two families of $N$-qubit states}

\author{Sanchit Srivastava}
\email{sanchit.srivastava@uwaterloo.ca}
\affiliation{Institute for Quantum Computing, University of Waterloo, Waterloo, Ontario, Canada N2L 3G1}
\affiliation{Department of Physics and Astronomy, University of Waterloo, Waterloo, Ontario, Canada N2L 3G1}
\author{Shohini Ghose}
\affiliation{Institute for Quantum Computing, University of Waterloo, Waterloo, Ontario, Canada N2L 3G1}
\affiliation{Department of Physics and Astronomy, University of Waterloo, Waterloo, Ontario, Canada N2L 3G1}
\affiliation{Department of Physics and Computer Science, Wilfrid Laurier University, Waterloo, Ontario, Canada N2L 3C5}

\begin{abstract}
Entangled states of multiple qubits can violate Bell-type inequalities 
indicating nonlocal behavior of multiqubit quantum correlations. We analyze the relation between multipartite entanglement and genuine multipartite nonlocality, characterized by Svetlichny inequality violations, for two families of $N-$qubit states. We show that for the generalized GHZ family of states, Svetlichny inequality is not violated when the $n-$tangle is less than $1/2$ for any even number of qubits. On the other hand, the maximal slice states always violate the Svetlichny inequality when $n-$tangle is nonzero, and the violation increases monotonically with tangle. Our work generalizes the relations between tangle and Svetlichny inequality violations previously derived for three qubits.  
\end{abstract}

\maketitle

\section{Introduction}
 Bell's inequalities provide a practical method for testing whether correlations observed between spatially separated parts of a system are compatible with any local hidden variable (LHV) description \cite{bell_einstein_1964}. Quantum entangled states can violate Bell-type inequalities, thereby confirming the fundamentally different nature of quantum correlations compared to those allowed by LHV models \cite{clauser_proposed_1969, brunner_bell_2014}. 
For the case of 2-qubit pure states, entanglement and nonlocality, as measured by Bell inequality violations, are directly related \cite{gisin_bells_1991, popescu_generic_1992}\textemdash the more entanglement there is, the larger the violation. For multiqubit pure states, the relationship between nonlocality and entanglement is much more complicated \cite{zukowski_all_2002, augusiak_entanglement_2015} and has not been explored in much detail.
In recent years, multipartite nonlocality has been identified as a useful resource for various tasks such as the study of many-body systems \cite{frerot_probing_2023}, quantum communication and cryptography  \cite{scarani_quantum_2001, xiang_multipartite_2023, moreno_device-independent_2020, gangopadhyay_controlled_2022}, self testing \cite{bharti_graph-theoretic_2022, murta_self-testing_2023}, and a wide array of device-independent applications including detection and quantification of entanglement \cite{brunner_testing_2012, moroder_device-independent_2013, bhattacharya_improvement_2017, holz_genuine_2020, bancal_device-independent_2011}. Multipartite nonlocality has also been shown to be relevant to quantum steering \cite{he_genuine_2013}, and the study of indefinite causal order in quantum mechanics \cite{curchod_multipartite_2014, abbott_genuinely_2017, baumeler_unlimited_2022}. 

 In this letter, we generalize the well-known relationship between 2-qubit entanglement and violation of the Bell-CHSH inequality \cite{clauser_proposed_1969} to the case of $N$-qubit entanglement and violation of an $N$-qubit Bell inequality.
For two qubits, all pure states can be written in the form $\ket{\psi} = \alpha \ket{00} + \beta  \ket{11}$ in some basis via the Schmidt decomposition. The maximum expectation value of the Bell-CHSH operator with respect to these states is $\max[S_2]=2\sqrt{2}$ \cite{cirelson_quantum_1980}, which violates the LHV bound of $S_2 \leq 2$ if $\ket{\psi}$ is an entangled state ($\alpha, \beta \neq 0$). Here, we analyse nonlocal correlations in two subsets of the $N-$qubit GHZ class \cite{dur_three_2000} of states: the generalized  GHZ (GGHZ) states $\ket{\psi_g}$ and the maximal slice  (MS) states $\ket{\psi_s}$ \cite{carteret_local_2000}: 
\begin{equation}\label{GGHZ}
    \ket{\psi_g}= \cos \alpha \ket{0}^{\otimes N} + \sin \alpha \ket{1}^{\otimes N},    
\end{equation}
\begin{eqnarray}
    \ket{\psi_s} = \frac{1}{\sqrt{2}}\left[ \ket{0}^{\otimes N} + \ket{1}^{\otimes N-1}\left( \cos\alpha \ket{0} + \sin \alpha \ket{1} \right)\right]
\end{eqnarray}.
These families of states are generalizations of the well-known 3-qubit GHZ state $(N=3, \alpha =\pi/4)$, which is a useful entanglement resource for a variety of information processing protocols including dense coding and teleportation \cite{greenberger_going_2007, bouwmeester_observation_1999, roos_control_2004, ekert_nmr_1998}. $N-$qubit generalizations of the GHZ states are thus natural candidates to explore for extending these protocols to more than three qubits. 
Tracing out any qubit in the GGHZ and MS states leaves the remaining qubits in a mixed state (for all $\alpha > 0$), indicating genuine $N$-qubit entanglement. 

In order to examine the multiqubit nonlocal properties of these states, we consider the expectation value of the Svetlichny operator $S_N$ that generalizes the 2-qubit Bell-CHSH operator to the case of $N$ qubits \cite{svetlichny_distinguishing_1987, seevinck_bell-type_2002, collins_bell-type_2002}. Violation of the inequality $|\left< S_N \right>|\leq 2^{N-1}$ implies genuine $N$-qubit non-separability as opposed to partial nonlocal correlations between less than $N$ qubits. Determining the maximum value of $|\langle S_N \rangle|$ for arbitrary quantum states is analytically and computationally demanding, since the number of terms in $S_N$ grows exponentially with $N$ \cite{batle_computing_2016, xiao_tight_2024}. Recently, \cite{xiao_tight_2024} introduced an approach that reduces the problem to evaluating the largest singular value of the state’s correlation matrix. For the specific families of GGHZ and MS states, however, their inherent symmetries enable us to derive closed-form expressions for the maximum of $|\langle S_N \rangle|$. To investigate the relationship between multipartite entanglement and nonlocality, we express these maximum expectation values of the Svetlichny operator in terms of the $n$-tangle, a measure that quantifies genuine multipartite entanglement in such states \cite{coffman_distributed_2000, osborne_general_2006, wong_potential_2001}.

For the GGHZ states with even number of qubits ($N>2$), we show that the maximum of $S_N$ is 
\begin{equation}\label{eveng}
S_{N \max}^{\pm}(\psi_g)=\left\{\begin{array}{ll}
2^{\frac{N}{2}}, &  \tau(\psi_g) \leq 2^{1-N}\\
 2^{N-1} \sqrt{2\tau}, &  \tau(\psi_g) \geq 2^{1-N}. 
\end{array}\right.
\end{equation} 
Here, $\tau(\psi_g)$ is the $n$-tangle of the GGHZ state. 
Our analytical expressions show interesting features of the nonlocality of these states.
Unlike the 2-qubit case,  nonlocality does not increase monotonically with entanglement.
Most significantly, the inequality $|\left< S_N \right>|\leq 2^{N-1}$ is not violated when $\tau(\psi_g) \leq 1/2$, even though the states are genuinely $N$-qubit entangled. Thus, we show that the Svetlichny inequality is not sensitive to the $N$-qubit entanglement in these states. Interestingly, the critical value of $\tau(\psi_g) \leq 1/2$ beyond which violation occurs, is the same for all values of $N$.

For MS states with even number of qubits ($N > 2$), we derive the maximum of $S_N$ to be 
\begin{eqnarray}\label{evens}
    S_{N\max}^{\pm}(\psi_s) = 2^{N-2}\sqrt{1 + \tau(\psi_s)}
\end{eqnarray}
Where $\tau(\psi_s)$ is the $n-$tangle of the MS state. This result is identical to the one derived for the 3-qubit case in \cite{ghose_tripartite_2009, ajoy_svetlichnys_2010} in terms of the $3$-tangle.  
We see that, in contrast to the GGHZ states, the bound $|\left< S_N \right>| \le 2^{N-1}$ is always violated for MS states as long as the state is not bi-separable, i.e. $\tau(\psi_s) \neq 0$.

Our results indicate that the genuine multiqubit nonlocality of MS states make them potentially useful resources in quantum information processing applications \cite{moreno_device-independent_2020}. 
For the GGHZ states, our expressions can be used to confirm a previous conjecture: above a critical value of entanglement $\tau$, the $N-$qubit nonlocal correlations cannot be reproduced by allowing parties to join together or broadcast their measurement inputs \cite{bancal_quantifying_2009}. Furthermore, since these violations are achieved by local measurements on each qubit, our results can provide a practical approach for calculating the entanglement in these states experimentally.  

\section{Svetlichny Inequality}
Consider $N$ spatially separated particles and two dichotomic observables $A^0_{i},A^1_{i}$, $i=1,2\dots N$ for each particle. Then the following operator can be constructed: 
\begin{eqnarray}
S_N^{\pm} = \sum_{\Vec{x}} \nu^{\pm}(\Vec{x})A(\Vec{x})  
\end{eqnarray}
where $\{\Vec{x}\}$ are bit strings of size $N$, $A(\Vec{x})$ is an $N$-qubit operator of the form $   A(\vec{x}) = \bigotimes_{i=1}^{N} A_{i}^{x(i)}$ and $\nu^{\pm}(\vec{x})$ is a constant dependent on the Hamming weight $w(\vec{x})$ of the bit string
\begin{eqnarray}
    \nu^{\pm}(\vec{x}) = -1^{w(\vec{x})(w(\vec{x})\pm 1)/2}.
\end{eqnarray}
For $N=2$, the above expression reduces to the standard CHSH form \cite{clauser_proposed_1969} and for $N=3$, it yields the operator in \cite{svetlichny_distinguishing_1987}. Svetlichny showed that if one allows nonlocal correlation between at most $N-1$ of the parties, then $|\braket{S_N^{\pm}}| \leq 2^{N-1}.$ Violation of this inequality confirms genuine $N$-partite nonlocality. 

\section{Maximum violation for GGHZ states}
The two  dichotomic observables $A^0_{i},A^1_{i}$, $i=1,2\dots N$ for each qubit can be written in terms of the Pauli operators $\sigma_x,\sigma_y$ and $\sigma_z$ as $A^{x(i)}_{i} = \vec{v}^{x(i)}_{i}\cdot \Vec{\sigma}$, where $\vec{v}^{x(i)}_{i}$ are unit vectors in $\mathbbm{R}^3$ and $\vec{\sigma} = (\sigma_x,\sigma_y,\sigma_z)$.  Defining $\Vec{v}^{x(i)}_{i} = (\sin \theta^{x(i)}_{i} \cos \phi^{x(i)}_{i},\sin \theta^{x(i)}_{i} \sin \phi^{x(i)}_{i}, \cos \theta^{x(i)}_{i} ),$ the  expectation value of the Svetlichny operator for GGHZ states is
 \begin{eqnarray}
     \nonumber &&|\braket{S_N^{\pm}}| = \Big| c_1 \sum_{\vec{x}}\nu^{\pm}(\vec{x})\prod_{i}^{N}\cos \theta_{i}^{x(i)}\\ 
      &&+ c_2 \sum_{\vec{x}}\nu^{\pm}(\vec{x})\cos\left( \sum_{i}^{N} \phi_{i}^{x(i)} \right)\prod_{i}^{N} \sin \theta_{i}^{x(i)} \Big|,
 \end{eqnarray}
which can be written in the form $  |\braket{S_N^{\pm}}| = |c_1 F_N^{\pm} + c_2 G_N^{\pm}|,$ where $F_N$ and $G_N$ contain all the $\cos \theta$ and $\sin \theta$ terms of $S_N$ respectively and $c_1 = \cos^2 \alpha + (-1)^N \sin ^2 \alpha$, while $c_2 = \sin 2 \alpha$. We calculate the maximum value of $ |\braket{S_N^{\pm}}|$ by considering the following cases: 

When $\alpha =0$ or $\pi/2$, $c_2 =0$. The state becomes a product state $\ket{0}^{ \otimes N}$ or $\ket{1}^{ \otimes N}$ and the maximum expectation value corresponds to the classical bound for local hidden variable models: 
\begin{equation}\label{separable1}
\begin{aligned} 
|\braket{S_{N}^{\pm}}|_{\max}=\left|F_{N}^{\pm}\right|_{\max } &=\left|\sum_{k=0}^{N} (-1)^{k(k\pm1)}{N \choose k}\right|. 
\end{aligned}
\end{equation}

Evaluating the above sum,  we get
\begin{eqnarray}
    |\braket{S_{N}^{\pm}}|_{\max}=\left|F_{N}^{\pm}\right|_{\max } &=\left\{\begin{array}{ll}
2^{\frac{N+1}{2}}, &  N \text { is odd } \\
2^{\frac{N}{2}}, &  N \text { is even }.
 \end{array}\right.
\end{eqnarray}
This maximum is achieved by measuring all the qubits along the $Z$-axis (i.e., setting all $\theta_i^{x(i)}$ to $0$ or $\pi$). 

When  $\alpha= \pi/4$, we get the maximally entangled N-qubit GHZ state. Let us first consider the case of odd $N$. In this case the maximum expectation value is
\begin{eqnarray}\label{maximal1}
       \nonumber  &&|\braket{S_N^{\pm}}|_{\max} = |G^{\pm}_{N}|_{\max} =\\
         &&\max\left( \left| \sum_{\vec{x}}\nu^{\pm}(\vec{x})\cos\left( \sum_{i}^{N} \phi_{i}^{x(i)} \right)\prod_{i}^{N} \sin \theta_{i}^{x(i)} \right| \right).
\end{eqnarray}
Note that the function $\nu^{\pm}(\vec{x})$ can be written as $ \nu^{\pm}(\vec{x}) = \sqrt{2}\cos \left( \pm\frac{\pi}{4} \pm w(\vec{x})\frac{\pi}{2} \right)$ \cite{seevinck_bell-type_2002}. Hence, if we fix $\phi_1^0 = \phi_{1}^{1} = \pm \pi/4$, $\phi_i^0 = 0$ and $\phi_i^1 = \pi/2$ for all $i\neq 1$, then we see that the cosine term in Eq.(\ref{maximal1}) becomes equal to $\nu^{\pm}(\vec{x})$ and hence every term inside the summation becomes positive. Now setting all $\theta_i^{x(i)}=0$, we get the maxima 
\begin{eqnarray}\label{maximal}
         |\braket{S_N^{\pm}}|_{\max} = |G^{\pm}_{N}|_{\max} = \sqrt{2}2^{N-1}
\end{eqnarray}
which achieves the maximum value possible for any quantum state \cite{collins_bell-type_2002}. For the case of even $N$, the maximum expectation value is $ |\braket{S_N^{\pm}}|_{\max} = |F_N^{\pm} + G_N^{\pm}|_{\max} $. Notice that for the choice of angles for which $|G^{\pm}_N|$ is maximum, $|F^{\pm}_{N}| = 0$. Since this choice of angles achieves the maximum possible quantum bound, the maximum for the case of even $N$ is the same as the odd $N$ case. 

We can now derive the maxima for the general case of an arbitrary value of $\alpha$ using the above two cases. When $F_{N}^{\pm}$ is maximum $G_N^{\pm} = 0$ for the same choice of angles and vice versa. Consider the derivative of $\left< S_N^{\pm} \right>$ with respect to $\theta_k^l$ ($l = 0$ or 1):
\begin{eqnarray}\label{derivative}
   \nonumber \frac{\partial \left< S_N^{\pm} \right> }{\partial \theta_k^l } =&&  \sum_{\vec{x}(k) =l}\nu^{\pm}(\vec{x})\Big[-c_1 \sin \theta_k^l\prod_{i\neq k}^{N}\cos \theta_{i}^{x(i)}\\ +&& c_2\cos\left( \sum_{i}^{N} \phi_{i}^{x(i)} \right)\cos \theta_k^l\prod_{i \neq k}^{N} \sin \theta_{i}^{x(i)}\Big].  
\end{eqnarray}
We see that $\partial  \left< S_N^{\pm} \right> / \partial \theta_k^l =0$ for the choices of $\{(\theta_i,\phi_i)\}$ where $F^{\pm}_N$ is maximum and for the choices where $G^{\pm}_N$ is maximum. It can be easily verified that the second derivative with respect to $\theta_k^l$ is negative at these points. Since Eq.(\ref{derivative}) is the form of the derivative with respect to all $\theta$, we can conclude that the points where $F^{\pm}_N$ and $G^{\pm}_N$ are maximum are local maxima for $\left< S_N^{\pm} \right>$. As the first term $(c_1F^{\pm}_N)$ is independent of $\phi$, checking the derivative with respect to $\phi$ here is not required. Considering the larger one of these two maxima, we get
\begin{eqnarray}\label{local}
   \nonumber |\braket{S_N^{\pm}}|^{\text{local}}_{\max} = \max\Big[&&\cos^2 \alpha + (-1)^N \sin ^2 \alpha|F_N^{\pm}|_{\max}, \\ 
    &&\sin2 \alpha|G_N^{\pm}|_{\max}\Big]. 
\end{eqnarray}
We find that the maximum achieved with this approach is, in fact, the global maxima. The proof for this is provided in Appendix \ref{appendixA}. Hence the maximum value in case of odd $N$ is
\begin{eqnarray}\label{odd}
\braket{S_N^{\pm}}_{\max}=\left\{\begin{array}{ll}
2^{\frac{N+1}{2}} \cos (2 \alpha), &  |\tan (2 \alpha)| \leq 2^{1-\frac{N}{2}} \\
\sqrt{2} 2^{N-1} \sin (2 \alpha), &   |\tan (2 \alpha)| \geq 2^{1-\frac{N}{2}} 
\end{array}\right.
\end{eqnarray}
and for the case of even $N$ is 
\begin{eqnarray}\label{even}
\braket{S_N^{\pm}}_{\max}=\left\{\begin{array}{ll}
2^{\frac{N}{2}}, & \sin (2 \alpha) \leq  2^{\frac{1-N}{2}} \\
\sqrt{2} 2^{N-1} \sin (2 \alpha), &  \sin (2 \alpha) \geq  2^{\frac{1-N}{2}}. 
\end{array}\right.
\end{eqnarray}
For the GGHZ state $\psi_g$ with even number of qubits, the $n$-tangle $\tau(\psi_g)$  is $\sin^2 2 \alpha$ \cite{wong_potential_2001}. Restating the above result in terms of $\tau(\psi)$ gives us the result in Eq.(\ref{eveng}). We see that our results match exactly the bounds derived in \cite{ghose_tripartite_2009, ajoy_svetlichnys_2010} for the case of $N=3$.

\section{Maximum violation for MS states}
To find the maximum of $S^{\pm}_N$ for the maximal slice states, we define unit vectors $\vec{b}^0$ and $\vec{b}^1$ such that $\vec{v}_{N-1}^0+\vec{v}_{N-1}^1 = 2\cos\theta \vec{b}^0$  and $\vec{v}_{N-1}^0-\vec{v}_{N-1}^1 = 2\sin\theta \vec{b}^1$. Thus 
\begin{eqnarray}\label{constraint}
   \nonumber\vec{b}^0\cdot\vec{b}^1 = &&\cos \theta_{b}^0 \cos \theta_{b}^1 \\
   &&+ \sin \theta_{b}^0 \sin \theta_{b}^1 \cos\left( \phi_b^{0} - \phi_b^{1} \right) = 0. 
\end{eqnarray}
By isolating the last two qubits, the Svetlichny operator can be written as 
\begin{eqnarray}
    \nonumber &&S^{\pm}_N = \sum_{\vec{\tilde{x}}}\nu^{\pm}(\vec{\tilde{x}}) A(\vec{\tilde{x}}) \left( A_{N-1}^0 + (-1)^{w(\vec{\tilde{x}})}A_{N-1}^1 \right)A_{N}^0\\
   &&+ \sum_{\vec{\tilde{x}}}\nu^{\pm}(\vec{\tilde{x}})A(\vec{\tilde{x}}) \left( (-1)^{w(\vec{\tilde{x}})}A_{N-1}^0 - A_{N-1}^1 \right)A_{N}^1 
\end{eqnarray}
where $\vec{\tilde{x}}$ are now bit strings of size $N-2$. Defining $B_{N-1}^k = \vec{b^k}\cdot \vec{\sigma}$ and using the fact 
\begin{equation}\label{trick}
    x\cos \theta + y\sin\theta \leq (x^2 + y^2)^{\frac{1}{2}},
\end{equation}
we get the inequality
\begin{eqnarray}\label{split}
   &&\nonumber\left< S^{\pm}_N \right> \leq2\sum_{\vec{\tilde{x}_e}}\left[ \left< A(\vec{\tilde{x}_e})B_{N-1}^0A_{N}^0 \right>^2 +  \left< A(\vec{\tilde{x}_e})B_{N-1}^1A_{N}^1 \right>^2 \right]^{\frac{1}{2}}\\ 
   && + 2\sum_{\vec{\tilde{x}_o}}\left[ \left< A(\vec{\tilde{x}_o})B_{N-1}^1A_{N}^0 \right>^2 +  \left< A(\vec{\tilde{x}_o})B_{N-1}^0A_{N}^1 \right>^2 \right]^{\frac{1}{2}}. 
\end{eqnarray}
Where $\{\vec{\tilde{x}_e}\}$ are bit strings of size $N-2$ and even weight,  and $\{\vec{\tilde{x}_o}\}$ are bit strings of size $N-2$ and odd weight. Consider the expectation value of the operator $A(\vec{\tilde{x}})B_{N-1}^kA_{N}^l$ with respect to $\psi_s$ for odd $N$:
\begin{widetext}
    \begin{eqnarray}\label{msterm}
     \nonumber\left< \psi_s| A(\vec{\tilde{x}})B_{N-1}^kA_{N}^l|\psi_s\right> &&\leq \Big(\cos^2\alpha\prod_{i =1}^{N-2} \cos^2 \theta_{i}^{\tilde{x}(i)} \left( \cos^2\alpha + \sin^2 \alpha \cos^2 (\phi_N^l) \right)\\
      &&+ \prod_{i =1}^{N-2} \sin^2 \theta_{i}^{\tilde{x}(i)}\left(  \cos^2 \alpha  \cos^2 (\phi(\tilde{x})+ \phi_{b}^k) + \sin^2 \alpha \cos^2 (\phi(\tilde{x})+ \phi_{b}^k + \phi_N^l) \right)\Big)^{\frac{1}{2}}
\end{eqnarray}
\end{widetext}
where $\phi(\tilde{x}) = \sum_{i}^{N-2}\phi_i^{\tilde{x}(i)}$ and we have used Eq.(\ref{trick}) for the angles $\theta_N^l$ and $\theta_b^k$ to obtain the inequality. The right hand side of Eq.(\ref{msterm}) can be maximized by setting all $\theta_i^{\tilde{x}(i)} =\pi/2$ . Hence we get 
    \begin{eqnarray}\label{msterm1}
     \nonumber \left< \psi_s|A(\vec{\tilde{x}})B_{N-1}^kA_{N}^l|\psi_s\right> && \leq \big( \cos^2 \alpha  \cos^2 (\phi(\tilde{x})+ \phi_{b}^k)\\
    +  \sin^2 \alpha \cos^2 &&(\phi(\tilde{x})+ \phi_{b}^k + \theta_N^l) \big)^{\frac{1}{2}}.
\end{eqnarray}

Inserting Eq.(\ref{msterm1}) into Eq.(\ref{split}) we obtain 

\begin{widetext}
\begin{eqnarray}\label{mspen}
  \nonumber\left< \psi_s|S_N^{\pm}|\psi_s \right>\leq 2 \sum_{\vec{\tilde{x}_e}}\Big[\cos^2 \alpha  \cos^2 (\phi(\tilde{x}_e)+ \phi_{b}^0) +&& \sin^2 \alpha \cos^2 (\phi(\tilde{x}_e)+ \phi_{b}^0 + \theta_N^0)\\
  \nonumber && + \cos^2 \alpha  \cos^2 (\phi(\tilde{x}_e)+ \phi_{b}^1) + \sin^2 \alpha \cos^2 (\phi(\tilde{x}_e)+ \phi_{b}^1 + \theta_N^1) \Big]^{\frac{1}{2}}\\
  \nonumber+ 2 \sum_{\vec{\tilde{x}_o}}\Big[\cos^2 \alpha  \cos^2 (\phi(\tilde{x}_o)+ \phi_{b}^1) +&& \sin^2 \alpha \cos^2 (\phi(\tilde{x}_o)+ \phi_{b}^1 + \theta_N^0)\\
  && + \cos^2 \alpha  \cos^2 (\phi(\tilde{x}_o)+ \phi_{b}^0) + \sin^2 \alpha \cos^2 (\phi(\tilde{x}_o)+ \phi_{b}^0 + \theta_N^1) \Big]^{\frac{1}{2}}
\end{eqnarray}
\end{widetext}

In Eq.(\ref{mspen}) the right had side can now be maximized by setting $\cos^2(\phi(\tilde{x})+ \phi_{b}^k + \phi_N^l) = 1$ $\forall \vec{\tilde{x}}, k, l$. This must be done by choosing angles such that $\phi_b^0-\phi_b^1 = \pi/2$ to satisfy the constraint Eq.(\ref{constraint}). Setting $\cos^2 (\phi(\tilde{x})+ \phi_{b}^0) = \sin^2 (\phi(\tilde{x})+ \phi_{b}^1)$ in Eq.(\ref{mspen}) we obtain 

\begin{eqnarray}
    \nonumber \left< \psi_s|S_N^{\pm}|\psi_s \right>&\leq& 2\sum_{\vec{\tilde{x}}}\left[ \cos^2 \alpha + 2\sin^2 \alpha \right]^{\frac{1}{2}}\\ 
    &=& 2^{N-1}\sqrt{1+\sin^2 \alpha}
\end{eqnarray}
since there are $2^{N-2}$ such bit strings. A similar analysis follows for the case of even $N$, giving the same bound. For the MS states $\psi_s$ with even number of qubits, substituting $\sin^2 \alpha = \tau(\psi_s)$ yields the result given in Eq.(\ref{evens}).

\section{Discussion}
While some previous studies have used numerical calculations to explore multiqubit nonlocality \cite{bancal_quantifying_2009, scarani_spectral_2001}, no analytical expressions for N-qubit Svetlichny inequality violations have been calculated to our knowledge. Here, we have analytically derived the maximum violation of the N-qubit Svetlichny inequality for the GGHZ and MS states for any number of qubits. This also provides an explicit relationship between multipartite entanglement and genuine multipartite nonlocality for these families of states when the number of qubits is even. Our results show the counterintuitive relationship between multipartite entanglement and nonlocality, and also highlight the very different behavior of these quantities for different entangled states. Although both GGHZ states and MS states are multiqubit-entangled, not all entangled GGHZ states violate the Svetlichny inequality.

Unlike the GGHZ states, all multipartite-entangled MS states violate the inequality. 
Furthermore, the entanglement-nonlocality relationship for MS states is identical to the well-known relationship between entanglement and nonlocality in 2-qubit pure states \cite{gisin_bells_1991}. 
Interestingly, in GGHZ states, the range of $n$-tangle values for which the Svetlichny inequality is not violated $(\tau(\psi_g) < 1/2)$ is independent of the number of qubits. 

Tangle as a measure of entanglement quantifies the genuine multipartite entanglement in a state,  
taking into account entanglement  not reducible to bipartite correlations \cite{coffman_distributed_2000, miyake_classification_2003}.
Some insight about the difference in the behavior of the Svetlichny inequality violation for GGHZ and MS states can be obtained from considering the tangle and hence the structure of entanglement distribution in these states. In GGHZ states, entanglement arises from a coherent superposition of the $\ket{0}^{\otimes N}$ and $\ket{1}^{\otimes N}$ components, and significant interference between these components is required to generate genuine multipartite nonlocal correlations \textemdash hence the threshold behavior. 
In contrast, the MS states have a more delocalized and asymmetrically distributed entanglement, where even a small contribution from the $\ket{1}^{\otimes N}$ component is enough to generate the nonlocal correlations that violate the Svetlichny inequality. This behavior shows that the geometry and the coherence structure of the entangled components, rather than the amount of entanglement alone, play a crucial role in determining the onset of genuine multipartite nonlocality. 

In addition to its applications in secret sharing, controlled teleportation, and key distribution protocols  \cite{moreno_device-independent_2020, gangopadhyay_controlled_2022, xiang_multipartite_2023}, the Svetlichny inequality is also useful as a strong form of self-testing \cite{murta_self-testing_2023}. Furthermore, in \cite{abbott_genuinely_2017}, the authors point out that the conditions imposed by the Svetlichny inequality can be used to define genuine multipartite non-causality, without the need for additional constraints.  
The identification of sharp thresholds in the relationship between multipartite entanglement and genuine multipartite nonlocality provides a concrete benchmark for quantum information tasks that rely on strong nonlocal correlations. 
Our finding that generalized GHZ states require a tangle exceeding $1/2$ to exhibit Svetlichny violation delineates the minimum entanglement cost for such protocols. Conversely, the monotonic violation observed for maximal-slice states with increasing tangle suggests that such states are particularly well-suited for nonlocality-based tasks in noisy or resource-limited settings. Our results thus provide important insights in the ongoing study of the precise operational roles of different types of multipartite entanglement within the framework of resource theories, particularly those concerned with nonlocality as a quantifiable and consumable resource.

 \section*{Acknowledgment}
This work was supported by the Natural Sciences and Engineering Research Council of Canada. Wilfrid Laurier University is located in the traditional territory of the Neutral, Anishnawbe and Haudenosaunee peoples. We thank them for allowing us to conduct this research on their land.

\bibliography{main}

\appendix
\section{Proof of global maximum}\label{appendixA}
For arbitrary values of $\alpha$, we proceed in the following way. We can single out, without loss of generality, the last two parties and label their measurement settings as $\vec{v}^{i}_{N-1},\vec{v}^{j}_{N}$, $i,j=0,1$. $S_N$ can now be written as 
\begin{eqnarray}
\label{isolate}
\nonumber\left|\left\langle S_N^{ \pm}\right\rangle\right|&& =\mid f_1^{ \pm} \cos \theta_{N-1}^0 \cos \theta_N^0+g_1^{ \pm} \sin \theta_{N-1}^0 \sin \theta_N^0 \\
\nonumber +f_2^{ \pm}&&\cos \theta_{N-1}^0 \cos \theta_N^1+g_2^{ \pm} \sin \theta_{N-1}^0 \sin \theta_N^1 \\
\nonumber+f_3^{ \pm}&&\cos \theta_{N-1}^1 \cos \theta_N^0+g_3^{ \pm} \sin \theta_{N-1}^1 \sin \theta_N^0 \\
 +f_4^{ \pm}&&\cos \theta_{N-1}^1 \cos \theta_N^1+g_4^{ \pm} \sin \theta_{N-1}^1 \sin \theta_N^1\mid 
\end{eqnarray}
where 
\begin{eqnarray}\label{15}
         \nonumber   f_1^{\pm} = c_1F_{N-2}^{\pm} \;&,&\; g_1^{\pm} = c_2\tilde{G}_{00,N-2}^{\pm}, \\ 
\nonumber  f_2^{\pm} = \pm c_1F_{N-2}^{\mp} \;&,&\; g_2^{\pm} = \pm c_2\tilde{G}_{01,N-2}^{\mp}, \\ 
\nonumber  f_3^{\pm} = \pm c_1F_{N-2}^{\mp} \;&,&\; g_3^{\pm} = \pm c_2\tilde{G}_{10,N-2}^{\mp}, \\ 
f_4^{\pm} = -c_1F_{N-2}^{\pm} \;&,&\; g_4^{\pm} = -c_2\tilde{G}_{11,N-2}^{\pm}.
\end{eqnarray}
Here, $\tilde{G}_{ij,N-2}^{\mp}$ have the same form as $G^{\pm}_{N-2}$ but with $\phi^{i}_{N-1} + \phi^{j}_N$ added to the argument of the cosine terms within the summation. 

From here on, we confine ourselves to $N>2$. Using the inequality in Eq.(\ref{trick}) for $\theta^{0}_N$ and $\theta^{1}_N$ in Eq.(\ref{isolate}) we get  
 \begin{eqnarray}\label{17}
               \nonumber|\braket{S_N^{\pm}}| \leq && \sqrt{f_1^{\pm^{2}}\cos^{2}\theta^{0}_{N-1} + g_1^{\pm^{2}}\sin^{2}\theta^{0}_{N-1} } \\ 
           \nonumber &+& \sqrt{f_2^{\pm^{2}}\cos^{2}\theta^{0}_{N-1} + g_2^{\pm^{2}}\sin^{2}\theta^{0}_{N-1} } \\
           \nonumber &+& \sqrt{f_3^{\pm^{2}}\cos^{2}\theta^{1}_{N-1} + g_3^{\pm^{2}}\sin^{2}\theta^{1}_{N-1} } \\
         &+& \sqrt{f_4^{\pm^{2}}\cos^{2}\theta^{1}_{N-1} + g_4^{\pm^{2}}\sin^{2}\theta^{1}_{N-1} } . 
 \end{eqnarray}

It is possible to check that the inequality in the last step is not a loose bound and that we can actually reach the equality for some choice of the variables. Inequality (\ref{trick}) turns into an equality when $\tan \theta = y/x$. Applying this to each of the four terms in Eq.(\ref{17}), we obtain the set of equations: 
\begin{equation}
\begin{array}{ll}
\tan \theta^{0}_N=\tan \theta^{0}_{N-1} \frac{g_{1}^{\pm}}{f_{1}^{\pm}} \quad, \quad \tan \theta^{0}_N=\tan \theta^{1}_{N-1} \frac{g_{2}^{\pm}}{f_{2}^{\pm}} \\ 
\\ \tan \theta^{1}_N=\tan \theta^{0}_{N-1} \frac{g_{3}^{\pm}}{f_{3}^{\pm}} \quad, \quad \tan \theta^{1}_N=\tan \theta^{1}_{N-1} \frac{g_{4}^{\pm}}{f_{4}^{\pm}}
\end{array}
\end{equation}

For the rest of our argument, at each stage, we make sure that these equations, which we call the consistency equations, are satisfied. 
Now define $H_N^{\pm}$ as the sum of the first two terms in Eq.(\ref{17}): 
\begin{eqnarray}
         \nonumber  H_N^{\pm} =  && \sqrt{f_1^{\pm^{2}}\cos^{2}\theta^{0}_{N-1} + g_1^{\pm^{2}}\sin^{2}\theta^{0}_{N-1} } \\ 
           &+& \sqrt{f_2^{\pm^{2}}\cos^{2}\theta^{0}_{N-1} + g_2^{\pm^{2}}\sin^{2}\theta^{0}_{N-1} }.
\end{eqnarray}
Setting $\partial H^{\pm}_{N}/\partial \theta^{0}_{N-1} =0$, we find that the maxima of $H^{\pm}_N$ occur at $\theta^{0}_{N-1} = 0, \pi/2$. When $\theta^{0}_{N-1} =0$, $ H_N^{\pm} = |f_1^{\pm}| + |f_2^{\pm}|$ and when $\theta^{0}_{N-1} =\pi/2$, $ H_N^{\pm} = |g_1^{\pm}| + |g_2^{\pm}|$. Therefore, the maximum of $ H_N^{\pm}$ is  
\begin{eqnarray}\label{hmax}
    \max[H_N] = \max(f_{\max},g_{\max})
\end{eqnarray}
where 
\begin{equation}
\begin{array}{l}
f_{\max }=\max \left[\left|f_{1}^{\pm}\right|+\left|f_{2}^{\pm}\right|\right] \\
g_{\max }=\max \left[\left|g_{1}^{\pm}\right|+\left|g_{2}^{\pm}\right|\right].
\end{array}
\end{equation}
Using Eq.(\ref{15}), 
\begin{eqnarray}
\nonumber f_{\max }&=&\max \left[c_{1}\left|F_{N-2}^{\pm} \pm F_{N-2}^{\mp}\right|\right]\\
&=&\max \left[c_{1}\left|F_{N-2}^{\pm}\right|\right]
\end{eqnarray}
where the second equality comes from the fact that $F^{\pm}_{N-2}$ and $F^{\mp}_{N-2}$ are symmetrically related in such a way that half the terms cancel out when the two are added together, yielding $\left|F_{N-2}^{\pm} \pm F_{N-2}^{\mp}\right|\leq \left|F_{N-2}^{\pm}\right| $. Using Eq.(\ref{separable1}), we get 
\begin{equation}\label{fmax}
f_{\max }=\left\{\begin{array}{ll}
c_{1} 2^{\frac{N-1}{2}} & \text { if } N \text { is odd } \\
c_{1} 2^{\frac{N-2}{2}} & \text { if } N \text { is even }.
\end{array}\right.
\end{equation}
Similarly, using Eq.(\ref{15}), 
\begin{eqnarray}
\nonumber g_{\max }&=&\max \left[c_{2}\left|\tilde{G}_{00, N-2}^{\pm} \pm \tilde{G}_{01, N-2}^{\mp}\right|\right]\\ 
&=&\max \left[2 c_{2}\left|G_{N-2}^{\pm}\right|\right]. 
\end{eqnarray}
In this case, the second equality has an additional factor of two because we can choose azimuthal angles $\phi_{(i)}^{x(i)}$ along with $\phi^{0}_{N}$ and $\phi^{1}_{N}$ such that both $\tilde{G}_{00,N-2}^{\mp}$ and $\tilde{G}_{01,N-2}^{\mp}$ achieve their maximum value concurrently. Using Eq.(\ref{maximal}), we get 
\begin{eqnarray}\label{gmax}
          g_{\max} = c_2\sqrt{2}2^{N-2}. 
\end{eqnarray}
Substituting Eq.(\ref{fmax}) and \ref{gmax} into Eq.(\ref{hmax}) we obtain:
For odd $N$ 
\begin{eqnarray}
\nonumber\max &&\left[H_{N}^{\pm}\right]_{\text {odd }}=\\
&&\left\{\begin{array}{ll}
2^{\frac{N-1}{2}} \cos (2 \alpha) & \text { if } 2^{\frac{N}{2}} |\tan (2 \alpha)| \leq 2 \\
\sqrt{2} 2^{N-2} \sin (2 \alpha) & \text { if } 2^{\frac{N}{2}} |\tan (2 \alpha)| \geq 2.
\end{array}\right.
\end{eqnarray}
For even $N$ 
\begin{eqnarray}
\nonumber \max && \left[H_{N}^{\pm}\right]_{\text {even }}=\\
&&\left\{\begin{array}{ll}
2^{\frac{N-2}{2}} & \text { if } 2^{\frac{N}{2}} \sin (2 \alpha) \leq  \sqrt{2} \\
\sqrt{2} 2^{N-2} \sin (2 \alpha) & \text { if } 2^{\frac{N}{2}} \sin (2 \alpha) \geq  \sqrt{2}. 
\end{array}\right.
\end{eqnarray}

We can follow the same procedure as above for the last two terms of Eq.(\ref{17}) (denote their sum as $K^{\pm}_{N}$), this time maximizing over $\theta^{1}_{N-1}$ in the first step, obtaining the same maxima for $K^{\pm}_{N}$.
Since $\max[H^{\pm}_N] = \max[K^{\pm}_N]$, substituting these into Eq.\ref{17} we get the bound
\begin{eqnarray}\label{final}
     |\braket{S_N^{\pm}}| \leq 2\max[H^{\pm}_N] = 2\max[K^{\pm}_N]. 
\end{eqnarray}
In the above procedure, apart from the angles $\theta^{0}_{K-1}$ for $H^{\pm}_{N}$ and $\theta^{1}_{K-1}$ for $K^{\pm}_{N}$, the choice of angles that achieve the maximum for $H^{\pm}_N$ and $K^{\pm}_N$ are the same, so the maximum for both functions can be reached simultaneously. Hence, this bound is tight. Finally, inserting the maximum values of $H^{\pm}_N$ and $K^{\pm}_N$ into Eq.(\ref{final}), we arrive at the maximum expectation value of $S^{\pm}_N$ given in Eq.(\ref{odd}) and (\ref{even}). This is the same maxima achieved in Eq.(\ref{local}).

 \end{document}